\documentclass[prl,twocolumn,nofootinbib]{revtex4-1}

\usepackage{amsmath,amssymb, pxfonts}
\usepackage{graphicx}
\usepackage{ytableau}
\usepackage{pstricks}

\newcommand{\vev}[1]{\langle #1\rangle}
\newcommand{\ket}[1]{| #1\rangle}

\newcommand{\beq}{\begin{equation}}
\newcommand{\beqs}{\begin{equation*}}
\newcommand{\eeq}{\end{equation}}
\newcommand{\eeqs}{\end{equation*}}

\allowdisplaybreaks

\preprint{}

\begin{document}


\title{Topology and geometry  cannot be measured by an operator measurement in quantum gravity}

\author{David Berenstein} \author{Alexandra Miller}\affiliation{ Department of Physics, University of California, Santa Barbara, CA 93106}

\begin{abstract}In the context of LLM geometries, we show that superpositions of classical coherent states of trivial topology can give rise to new classical limits where the
topology of spacetime has changed. We argue that this phenomenon implies that neither the  topology nor the geometry of spacetime can be the result of an operator measurement. 
We address how to reconcile these statements with the usual semiclassical analysis of low energy effective field theory for gravity. 
\end{abstract}

\maketitle

One of the main claims of the AdS/CFT correspondence \cite{Maldacena:1997re} is that it provides a definition of quantum gravity for spacetimes that are asymptotically of the form $AdS\times X$. It is natural to ask:  what does this holographic description tell us about the nature of observables in the quantum theory of gravity? 

By an observable, we mean a Hermitian (linear) operator on the Hilbert space of states as is usual in quantum mechanics. In this context, 
 is the metric a quantum  mechanical observable?
 Is topology measurable by an observable?
 And if the answer is no, then when are they sufficiently well approximated by observables?

We define $\hat T$ to be a topology measuring operator if different eigenvalues correspond to different topologies
of the dual gravity theory and the zero eigenvalue is reserved for the trivial topology alone. Here trivial means the same topology as the ground state. Our main conclusion is that such topology measuring operators do not always exist. We support this by providing an example where one can prove that there is no such operator.

The example arises from studying the states that preserve half of the supersymmetries  of ${\cal N}=4 $ Super -Yang-Mills theory (SYM) and their dual geometries. 

The set of states we are interested in forms a Hilbert space in its own right. Quantum mechanics is therefore valid and quantum mechanical questions can be answered unambiguously. The relevant Hilbert space of states near the free field theory limit $g_{YM}\to 0$ has been analyzed in \cite{Corley:2001zk}. An orthogonal basis of states of energy $E=n$ can be represented by partitions of $n$, which can be written in terms of Schur polynomials and are classified by Young tableaux for $U(N)$. These states can also be represented in terms of free fermion dynamics for $N$ fermions in the lowest Landau level  on a plane \cite{Berenstein:2004kk}. This description gives rise to a geometric interpretation of states as incompressible droplets in two dimensions. These free fermions can also be described by the incompressible droplets of the integer quantum Hall effect \cite{QHE}.

The geometric droplet shape is exactly the geometric data that is required to build a horizon-free solution of type IIB supergravity that respects the same amount of supersymmetry and that also asymptotes to $AdS_5\times S^5$, as constructed by Lin, Lunin, and Maldacena \cite{Lin:2004nb}. We will call these the LLM geometries. In these geometries, different droplet topologies correspond to different spacetime topologies.

There exists a limit of the LLM geometries where a complete minisuperspace theory characterizing all the  states with the requisite amount of supersymmetry, as a quantum theory, is identical to the Hilbert space of a free chiral boson on a circle in 1+1 dimensions. This limit is the strict $N\to \infty$ limit of the theory, with the energy above the ground state kept finite. The mode expansion of the chiral boson can be related to traces of the ${\cal N}=4 $ SYM fields $Z$ by $a^\dagger_n\simeq \hbox{tr} (Z^n)$ via the usual operator-state correspondence and the understanding that single traces go to single particle modes \cite{Witten:1998qj}. In this limit, the oscillators give a rise to a free Fock space, with a free mode for every $n$. We will take the existence of this limit as a statement of fact and it is in this limit that our statements can be made rigorously.  Many of the technical details that are required to prove some claims in this paper will appear in a forthcoming paper by the authors \cite{We}.

This paper makes the claim that topology  cannot be measured by operators. To make the claim, we need the following assumptions about the particular setup we have:

\begin{enumerate}
\item All coherent states of the chiral boson theory with finite energy have trivial topology (the same as the vacuum) and are to be thought of as smooth classical geometries.
\item The set of these coherent states is over-complete, so every other state in the Hilbert space can be obtained by superposition of this family of states.
\item There are states in the Hilbert space that have a different topology than the vacuum and can also be thought of as classical states of the gravitational theory.
\end{enumerate}

From these assumptions, it follows that there is no operator $\hat T$ in the Hilbert space that measures the topology.  We now prove this statement by contradiction, assuming the existence of $\hat T$.

From assumption one above, all coherent states have trivial topology, so $\hat T \ket{\rm{Coh}} = 0$. Any other state $\ket \psi$ that is a superposition of coherent states will satisfy 
\begin{equation}
\hat T \ket \psi = \hat T\int_{\rm{Coh}} A_{\rm{Coh}} \ket {\rm{Coh}}= \int_{\rm{Coh}} A_{\rm{Coh}}  \hat T \ket {\rm{Coh}}=0
\end{equation}
so the ket $\ket \psi$ is an eigenstate of the topology operator with eigenvalue zero: it has trivial topology.
By condition two above, this includes all possible states. Therefore, if such an operator exists, all states have trivial topology. This contradicts the third assumption A related argument where overcompleteness is used to indicate problems with defining either topology operators or geometric operators is found in \cite{Motl,Papadodimas:2015jra}. These arguments are made in the  $ER=EPR$ context \cite{Maldacena:2013xja} for setups with entangled black holes, and the topology change is hidden behind a horizon in an Einstein-Rosen bridge. 

We will now elaborate on the basis for assumptions one and three. Assumption two is a well known fact for studying states of a finite number of harmonic oscillators.  It can be extended to the case of an infinite number of oscillators by carefully taking the appropriate limits. 

A geometric picture of the states can be obtained as follows: in the LLM geometries, all states can be drawn as a two color picture in two dimensions. The individual droplet areas of both colors 
are quantized. As we are focusing on $N\to \infty$, keeping the energy finite, all relevant states are close to the circular droplet that makes the vacuum. We want to focus on the edge of the droplet, by using an area preserving map
\begin{equation}
d x \, dy \simeq r \, dr \, d\theta = dh \, d \theta 
\end{equation}
where the variable $h$ will be measured relative to the circular droplet. In this setup, the $N\to \
\infty$ limit is taken by  sending $r\to \infty$, keeping $h$ finite. In this limit, $|h|$ can be as large as we need it to be. The topology of the $(h, \theta)$ space is a cylinder. The vacuum has 
the area below $h=0$ completely filled, and above $h=0$ completely empty. We can excite fermions from the filled region to the empty region and will characterize this shift by a density function $\Delta\rho$, which takes on a value $+1$ for regions above $h=0$ and $-1$ below. Conservation of the fermion number is implemented by $\int dh \, d\theta \, \Delta \rho=0$. The energy relative to the vacuum is measured by
\begin{equation}
E \simeq \int d\theta \, dh \,  h(\theta) \, \Delta \rho(h, \theta) \label{eq:energy}
\end{equation}
This follows easily from the computations in \cite{Lin:2004nb}, being careful about subtracting the energy of the vacuum.

\begin{figure}[ht]
\begin{center}
\begin{pspicture}(4,4)
\pscustom{
\gsave
    \psline(1,0)(1,1.8)
    \pscurve[linewidth=2pt](1,1.8)(1.5,1.85)(2,2.2)(2.5,1.85)(3,1.8)
    \psline(3,1.8)(3,0)
    \psline(3,0)(1,0)
    \fill[fillstyle=solid,fillcolor=black]
\grestore
}
\psline(1,0)(1,4)
\psline(3,0)(3,4)
\pscurve[linewidth=2pt,linecolor=magenta](1,1.8)(1.5,1.85)(2,2.2)(2.5,1.85)(3,1.8)
\psline[linestyle=dashed](1,2)(3,2)
\rput(2,2.6){$h(\theta)$}
\end{pspicture}
\end{center}
\caption{A geometric fluctuation of the vacuum, characterized by $h(\theta)$}\label{fig:h}
\end{figure}
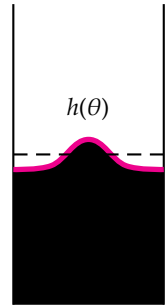

A typical geometric fluctuation is depicted in figure \ref{fig:h}. 

The fluctuation is described by a single height function $h(\theta)$ that represents the edge of the droplet.

The function $h(\theta)$ is the excess density of fermions at the angle $\theta$.  It gets matched to the charged current of the chiral boson as $h(\theta) \propto \partial_{\theta} X(\theta)$. Conservation of fermion number is described by $\int d\theta \, \partial_{\theta} X(\theta)=0$.  That is, the field $\partial_{\theta} X$ has no zero mode.  This is exactly as is expected from studies of the quantum Hall effect (see for example \cite{Stone:1990iw,Stone:1990ir}). It follows from integrating $\Delta\rho$ over a column in equation \eqref{eq:energy} that the energy goes to 
$E \simeq \frac 1{2\pi} \int d\theta :\partial_{\theta}X(\theta)^2:$, where the normal ordering ensures that the vacuum has zero energy. This is the standard expression for the energy in the chiral boson theory. The factor of $2\pi$ is a choice of convention for normalization of the field Fourier modes.

A coherent state of the free chiral boson will result in a unique (sufficiently smooth) single valued $h(\theta)\propto \vev{\partial X(\theta)}$ such that the classical energy of the state as computed in \eqref{eq:energy} is equal to the expectation value of the energy of the corresponding quantum state. All of these solutions have a classical LLM geometry that can be reconstructed uniquely from $h(\theta)$. The topology of the geometry is encoded in the topology of the fermion droplet. All of these states have trivial topology in the LLM setting: one edge with circle topology winding once around the circle direction $\theta$.
This justifies our assertion one made before.

\begin{figure}[ht]
\begin{center}
\includegraphics[width=2 cm]{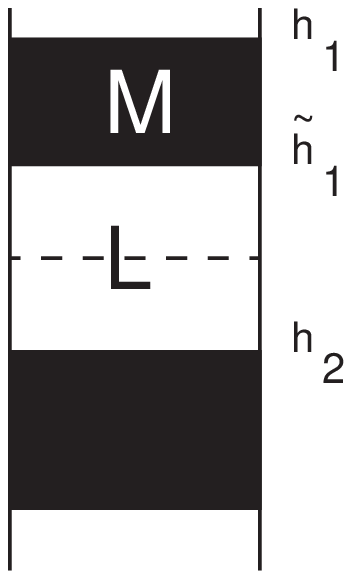}
\end{center}
\caption{A two coloring with non-trivial geometry. The areas $L,M$ have quantized are $L,M$ respectively.}\label{fig:multi-strip}
\end{figure}

Now we need to justify assertion three. This can be done with figure \eqref{fig:multi-strip}.
The idea is that we can also do a two coloring of the cylinder that preserves the net area and is such that the topology is now characterized by a strip-geometry. In this case, there are three edges winding around the circle, 
two of them go from black to white (at heights $h_1,h_2$) and the other one goes from white to black (at height $\tilde h_1$). These edges with the opposite coloring will be called anti-edges. We call this state the reference state $\ket{\square_{LM}}$. This state is easily constructible in terms of Young diagrams \cite{Lin:2004nb}.

Small fluctuations of this geometry will be characterized by three functions $h_1(\theta)= h_1^{(0)} +\delta h_1(\theta)$, $h_2(\theta)= h_2^{(0)} +\delta h_2(\theta)$, and $\tilde h_1(\theta)= \tilde h_1^{(0)} +\delta \tilde h_1(\theta)$. 
Quantization of the area is implemented by requiring that none of the $\delta h$ have a zero mode in the Fourier coefficients.  This can easily be generalized to more stripes. A straightforward computation of the energy of such a geometry shows that the energy, relative to the reference state, is given by
\begin{equation}
E \simeq E_{LM} +\int d\theta \sum_{i} \left( \delta h_i(\theta)^2-\delta \tilde h_i(\theta)^2\right) \label{eq:signed}
\end{equation}
with the edge modes having positive excess energy and the anti-edge modes having negative excess energy.
The net fermion over density at position $\theta$ is $\partial X(\theta) \simeq  h_1(\theta) + h_2(\theta) -\tilde h_1(\theta) $. The absence of the zero mode for $\partial X(\theta)$ results in $h_1^{(0)}+h_2^{(0)}-\tilde{h}_1^{(0)}=0$. This determines the location of the reference height, which tells us that the reference state depends only on $L,M$, with no extra parameters.

We now claim that the new topology is generated by making the height function $h(\theta)$ multivalued. This function is related linearly to $\partial X(\theta)$ in the classical coherent state setup. The net $\partial X(\theta)$ that reflects a proper observable  in the quantum system is obtained by a signed sum over these multi-values. Indeed, because all the edges are similar, one can imagine that to each of the edges one could associate a chiral boson field theory so that 
$\partial X(\theta)= \partial X_1(\theta)+\partial X_2(\theta)-\partial X_{\tilde 1}(\theta)$.  Because in equation \eqref{eq:signed} the tilded modes have the wrong sign, 
the notion of raising and lowering operators is reversed. We can rewrite this equation in a mode expansion
\begin{equation}
a_n^\dagger = b^{(1)\dagger}_n+b^{(2)\dagger}_n-c_n^{(\tilde 1)}\label{eq:linear}
\end{equation}
where the $b$ modes refer to regular edges, and the $c$ modes to the anti-edges. Notice, without the lowering operator pieces in equation \eqref{eq:linear}, the necessary commutation relations of the $a_n$ modes could not be satisfied. This also gives the correct equations of motion for $\partial X$, with each of the modes satisfying them on their own. The negative energy associated with the modes $c$ is crucial, so that the notion of positive and negative frequency can reverse the assignment of raising and lowering operators. This equation can be thought of as a partial Bogolubov transformation
mode by mode. The reference state is characterized by $b^{(i)}_n \ket{\square_{LM}}=c^{(\tilde i)}_n \ket{\square_{LM}}=0$ for all $n$.

The linearity of the mode decomposition for strip geometries has already been suggested in \cite{Koch:2008ah} (see also the more recent \cite{Koch:2016jnm}). The construction of such modes is purely combinatorial and depends on knowing how to manipulate the states labeled by Young tableaux  carefully.
The commutation relations of the $b,c$ modes are canonical for {\em states near the reference state}. This can be deduced from \cite{Grant:2005qc}. We take these to be
\begin{equation}
[b^{(i)}_n, b^{(j)\dagger}_n] = n \delta^{i,j}
\end{equation}
and similar for $c$, with all other commutators vanishing.
These assertions are  proven in the companion work to this paper \cite{We}, where the details on the cutoff and the applicability of these commutation relations are deduced from first principles. The nearby states form a small Hilbert space in their own right. The commutation relations are valid when inside the small Hilbert space, but they get corrected as we try to include more states.

These new modes only extend to values of order  $n<< M,L$. Beyond that they do not exist as independent operators \cite{We}. This is a type of stringy exclusion principle of the same type as the one implemented in \cite{McGreevy:2000cw}. It is dynamically generated and depends on the reference state (depends on $L,M$). The modes $b,c$ do not exist for any of the coherent states $\ket{\rm{Coh}}$
that we have discussed previously. For those states, the height function is single-valued. We should not be able to extend the definition of these modes to those states. We claim we are prevented from doing so by the stringy exclusion principle.  The existence of these states justifies our third assumption, and therefore completes our argument that one cannot have a topology measuring operator.

It seems we are lost.  Does this lack of a topology measuring operator mean we simply cannot determine the topology of the spacetime?  In the remainder of the paper, we will give two resolutions: one that involves measuring classicality of the state and one that involves its entanglement.  Both of these rely on computing quantities that are non-linear in the wavefunction, rather than performing a single operator measurement.

Consider forming coherent states of the $b,c$ oscillators, which can be interpreted as new classical solutions relative to the state $\ket{\square_{LM}}$, with $\delta h^i(\theta) \propto \vev{\partial X^i(\theta)}$ and similar for the anti-edges.  These are allowed as long as the tails in the coherent state can be truncated without appreciable loss of information.

The existence (construction) of the $b,c$ modes means we can do (unitary) effective field theory in the nearby Hilbert space  with them. We just need to restrict ourselves to being well below the stringy exclusion principle. The small Hilbert space is constructed by acting with finitely many raising operators $b^\dagger, c^\dagger$, keeping the total energy in the $b$ modes less than $\rm{min}(L/2,M/2)$, and the total negative energy in
the $c$ modes less than $\rm{min}(L/2,M/2)$.    In that regard, the operators $a_n^\dagger, a_n$ well below the (dynamical) stringy exclusion principle act inside the small Hilbert space, leaving the new state inside it.  Any quantum mechanical question about them can be answered in principle in the small Hilbert space: they belong to the effective field theory. This explains why effective field theory is still valid in the gravity theory.

Our first method comes from taking the expectation value of the number operator $\hat N_n$ for mode $a_n$  in the reference state (this is easy to do for multi-edge geometries).  We find that 
\begin{equation}
\vev {n^{-1}a_n^\dagger a_n }_{LM} = N_{\rm{anti-edges}}\quad \vev {n^{-1}a_n a_n^\dagger }_{LM} = N_{\rm{edges}}
\end{equation}
so the expectation value of the number operator (on reference states) can be used to measure the topology, mode per mode. The number operator can change a lot when we consider coherent states of the $b,c$ modes. Let us call one such state $\ket \psi$.
Consider instead of the number operator, the uncertainty. A straightforward manipulation shows that
\begin{equation}
\vev{n^{-1}( a_n^\dagger-\vev{a_n^\dagger}_\psi )(a_n-\vev{a_n}_\psi )}_\psi= N_{\rm{anti-edges}}
\end{equation}
We see the topology of the coherent state of the $b,c$ oscillators, the new classical states, can be measured by computing the net fluctuations of the modes $a_n^\dagger$. These are still of quantum size (order one), so the state can be said to be approximately classical for each of the modes $a_n$. In taking a double scaling limit $\hbar \to 0$, implemented by taking $L,M\to \infty$ and rescaling the fields by appropriate  powers of $L,M$, the rescaled uncertainty vanishes. In this sense, these topologically different drawings provide new classical limits of the free chiral boson.

The topology is measurable by the uncertainty. This is a non-linear operation in the Hilbert space: it is not a single operator measurement, but a test of classicality.

If we want to extend the measurement of topology to the semiclassical limit, where we  allow a few quanta of the $b,c$ modes to be in a state that is not a coherent state, we find that to measure the topology, we have to ask each mode $a_n$ what value of uncertainty they measure. The few modes that are outliers can be discarded and the majority rule will be used. We call this a census measurement. The best answer for the topology will be given by the consensus of the majority. This depends highly on what scale we use to cut off the census. This should be determined by the stringy exclusion principle, which is related to the value of $L,M$.  But, we do not know these a priori: the state is given to us as a black box.  If the cutoff is set at a scale much larger than $L,M$, most of the $a_n$ modes will be in the vacuum and we would find that state has a trivial topology. If the cutoff is set well below $L,M$, the consensus might give a different topology than if we measure near $L,M$. This is because the $b,c,$ modes may be forming thinner striped states on their own. Such a state is a geometry with a lot of bubbles and we should not necessarily associate it with a fixed semiclassical topology.

We will next use the idea that spacetime geometry and entanglement seem to be intimately related.  We compute the entanglement entropy using the Bogoliubov transformation.  Starting with a coherent state of the $b,c$ modes, we find
\begin{equation}
S_{n}=N_{\rm{edges}}\ln N_{\rm{edges}}-N_{\rm{anti-edges}}\ln N_{\rm{anti-edges}}
\end{equation}
where everything but the $a_n$ modes have been traced out.  As with the previous method, we need to perform this computation for many modes and find consensus to determine the topology of nearby semi-classical states.  And, again, we can only be sure of the accuracy of this calculation for modes below the stringy exclusion principle.  The connection we find between topology and entanglement supports the ideas of Van Raamsdonk \cite{VanRaamsdonk:2010pw}. Related ideas about connectedness being related to entanglement are currently being developed by Almheiri et al. \cite{Ahl}.

It is important to note that we have been working in the strict $N\rightarrow \infty$ limit.  At finite $N$, there is no longer a canonical factorization of the Hilbert space, so computing the entanglement entropy becomes  ambiguous.  This challenge suggests that the uncertainty measurement route might be preferable, where progress seems clearer.

Further,  at finite $N$, the Planck scale scales as $\ell_p^{-1}\simeq N^{1/4}$. If $L,M>>N^{1/4}$, there are many more modes with energy below $\ell_p^{-1}$ in the geometry with the striped topology than when computed in the ground state
of the system. These all commute with each other. 
To describe these multi-droplet geometries, one needs to borrow supersymmetric modes from the UV  \cite{Berenstein:2016mxt}.  To end up with the extra finite energy modes, whose energies are of order one, one needs the UV modes to be excited. That way, the UV modes don't annihilate the  reference state and one can form a bound state of a mode that raises the energy with another mode that lowers it. In the excited state, there is a finite probability that the lowering operators do not annihilate the excited state. These count as large negative energy excitations relative to the reference state.
Bound states at  threshold between the large positive  energy excitations and the large negative energy excitations provide a consistent solution to this quandary. This also provides explanation for the modes $c$ having negative energy. The precise details of such a construction will be taken up elsewhere \cite{We}.

In this paper, we have shown that states with non-trivial topology can be formed by superposing topologically trivial states and therefore find that topology cannot be determined by a single operator measurement.  We have  also proposed two methods for extracting the topology from a state, one based on uncertainty and the other based on entanglement. Both of these rely on computing quantities that are non-linear in the wavefunction.
 
{\em Acknowledgments:} 
We are very grateful to A. Almheiri, D. Gross, G. Horowitz,  D. Marolf, A. Puhm,  for many discussions. Work  supported in part by the department of Energy under grant {DE-SC} 0011702.

\end{document}